\documentclass[longauth]{aa}  

\usepackage{graphics,graphicx,graphbox}
\usepackage{txfonts}
\usepackage{natbib}
\usepackage{ulem}
\usepackage[flushleft]{threeparttable}
\usepackage{multirow}
\usepackage{gensymb}
\usepackage{scrextend}   
\usepackage[colorlinks=true, linkcolor=blue, citecolor=blue, draft=false]{hyperref}
\usepackage{xcolor}
\usepackage{orcidlink}

\begin{document} 
   \title{Noema formIng Cluster survEy (NICE): \\Discovery of a starbursting galaxy group with a radio-luminous core at $z$\,=\,3.95}
   \titlerunning{The NICE survey overview \&  LH-SBC3, a starbursting galaxy group at $z$\,=\,3.95}

\author{Luwenjia Zhou\inst{\ref{c1},\ref{c2}}\thanks{E-mail: \texttt{wenjia@nju.edu.cn}}\orcidlink{0000-0003-1687-9665}
\and Tao Wang\inst{\ref{c1},\ref{c2}} \thanks{E-mail: \texttt{taowang@nju.edu.cn}}
\and Emanuele Daddi\inst{\ref{c3}} 
\and Rosemary Coogan\inst{\ref{c3}} 
\and Hanwen Sun \inst{\ref{c1},\ref{c2}} 
\and Ke Xu \inst{\ref{c1},\ref{c2}} 
\and Vinodiran Arumugam \inst{\ref{c8}}
\and Shuowen Jin\inst{\ref{c4},\ref{c5}}
\and Daizhong Liu\inst{\ref{c6}} 
\and Shiying Lu\inst{\ref{c1},\ref{c2},\ref{c3}} 
\and Nikolaj Sillassen\inst{\ref{c4},\ref{c5}}
\and Yijun Wang\inst{\ref{c1},\ref{c2}}
\and Yong Shi\inst{\ref{c1},\ref{c2}}
\and Zhi-Yu Zhang\inst{\ref{c1},\ref{c2}}
\and Qinghua Tan\inst{\ref{c7}}
\and Qiusheng Gu\inst{\ref{c1},\ref{c2}}
\and David Elbaz \inst{\ref{c3}}
\and Aurelien Le Bail \inst{\ref{c3}}
\and Benjamin Magnelli \inst{\ref{c3}}
\and Carlos G{\'o}mez-Guijarro \inst{\ref{c3}}
\and Chiara d'Eugenio \inst{\ref{c-2}, \ref{c-3}}
\and Georgios E. Magdis \inst{\ref{c4},\ref{c5},\ref{c-1}}\orcidlink{0000-0002-4872-2294}
\and Francesco Valentino \inst{\ref{c-4}}\orcidlink{0000-0001-6477-4011}
\and Zhiyuan Ji \inst{\ref{c-5}}\orcidlink{0000-0001-7673-2257}
\and Raphael Gobat \inst{\ref{c-6}}
\and Ivan Delvecchio \inst{\ref{c-7}}
\and Mengyuan Xiao \inst{\ref{c-8}}\orcidlink{0000-0003-1207-5344}
\and Veronica Strazzullo \inst{\ref{c-9},\ref{c-99}}
\and Alexis Finoguenov \inst{\ref{c-10}}
\and Eva Schinnerer \inst{\ref{c6}}
\and R. Michael Rich \inst{\ref{c-11}}
\and Jiasheng Huang \inst{\ref{c21}}
\and Yu Dai \inst{\ref{c21}}
\and Yanmei Chen\inst{\ref{c1},\ref{c2}}
\and Fangyou Gao\inst{\ref{c1},\ref{c2}}
\and Tiancheng Yang\inst{\ref{c1},\ref{c2}}
\and Qiaoyang Hao\inst{\ref{c1},\ref{c2}}
}

\institute{School of Astronomy and Space Science, Nanjing University, Nanjing 210093, China \label{c1}
\and Key Laboratory of Modern Astronomy and Astrophysics (Nanjing University), Ministry of Education, Nanjing 210093, China \label{c2}
\and AIM, CEA, CNRS, Universit\'{e} Paris-Saclay, Universit\'{e} Paris Diderot, Sorbonne Paris Cit\'{e}, F-91191 Gif-sur-Yvette, France  \label{c3}
\and IRAM, 300 rue de la piscine, F-38406 Saint-Martin d'H\`{e}res, France\label{c8}
\and Cosmic Dawn Center (DAWN), Jagtvej 128, DK2200 Copenhagen N, Denmark \label{c4}
\and DTU-Space, Technical University of Denmark, Elektrovej 327, 2800 Kgs. Lyngby, Denmark \label{c5}
\and Max-Planck-Institut f\"{u}r Extraterrestrische Physik (MPE), Giessenbachstrasse 1, 85748 Garching, Germany \label{c6}
\and Purple Mountain Observatory, Chinese Academy of Sciences, 10 Yuanhua Road, Nanjing 210023, China \label{c7}
\and Niels Bohr Institute, University of Copenhagen, Jagtvej 128, DK-2200 Copenhagen N, Denmark\label{c-1}
\and Instituto de Astrofísica de Canarias, C. V\'{i}a L\'{a}ctea, s/n, 38205 La Laguna, Tenerife, Spain\label{c-2}
\and Universidad de La Laguna, Dpto. Astrofísica, 38206 La Laguna, Tenerife, Spain\label{c-3}
\and European Southern Observatory, Karl-Schwarzschild-Str. 2, D-85748 Garching bei Munchen, Germany\label{c-4}
\and Steward Observatory, University of Arizona, 933 N. Cherry Avenue, Tucson, AZ 85721, USA\label{c-5}
\and INAF - Osservatorio Astronomico di Brera, via Brera 28, I-20121, Milano, Italy \& via Bianchi 46, I-23807, Merate, Italy \label{c-7}
\and Department of Astronomy, University of Geneva, Chemin Pegasi 51, 1290 Versoix, Switzerland\label{c-8}
\and Instituto de F\'{i}sica, Pontificia Universidad Cat\'{o}lica de Valpara\'{i}so, Casilla 4059, Valpara\'{i}so, Chile\label{c-6}
\and INAF - Osservatorio Astronomico di Trieste, Via Tiepolo 11, 34131, Trieste, Italy \label{c-9}
\and IFPU - Institute for Fundamental Physics of the Universe, Via Beirut 2, 34014, Trieste \label{c-99}
\and  Department of Physics, University of Helsinki, Gustaf H\"{a}llstr\"{o}min katu 2, FI-00014 Helsinki, Finland\label{c-10}
\and Department of Physics \& Astronomy, University of California Los Angeles, 430 Portola Plaza, Los Angeles, CA 90095, USA\label{c-11}
\and Chinese Academy of Sciences South America Center for Astronomy (CASSACA), National Astronomical Observatories of China (NAOC), 20A Datun Road \label{c21}
}

\date{Received --; accepted --}

 
  \abstract{
 The study of distant galaxy groups and clusters at the peak epoch of star formation is limited by the lack of a statistically and homogeneously selected and spectroscopically confirmed sample.  Recent discoveries of concentrated starburst activities in cluster cores have opened a new window to hunt for these structures based on their integrated IR luminosities.  Hereby we carry out the large NOEMA (NOrthern Extended Millimeter Array) program targeting a statistical sample of infrared-luminous sources associated with overdensities of massive galaxies at $z$\,>\,2, the Noema formIng Cluster survEy (NICE).  We present the first result from the ongoing NICE survey, a compact group at $z$\,=\,3.95 in the Lockman Hole field (LH-SBC3), confirmed via  four massive ($M_{\star}$\,$\gtrsim$\,$10^{10.5}$\,M$_{\odot}$) galaxies  detected in CO(4-3) and [CI](1-0) lines. The four CO-detected members of LH-SBC3 are distributed over a 180 kpc physical scale, and the entire structure has an estimated halo mass of $\sim$\,$10^{13}$\,M$_{\odot}$ and total star formation rate (SFR) of $\sim$\,4000\,M$_{\odot}$/yr. In addition, the most massive galaxy hosts a radio-loud AGN with $L_{\rm 1.4GHz, rest}$\,=\,3.0\,$\times$\,10$^{25}$\,W\,Hz$^{-1}$. 
 The discovery of LH-SBC3 demonstrates the feasibility of our method to efficiently identify high-$z$ compact groups or forming cluster cores. The existence of these starbursting cluster cores up to $z$\,$\sim$\,$4$ provides critical insights into the mass assembly history of the central massive galaxies in clusters.  
  }


   \maketitle
%

\section{Introduction}

Galaxy groups and clusters at cosmic noon and earlier are  key to understanding the physical processes that lead to the emergence of the peak in the cosmic star formation history  \citep{Madau2014, Overzier2016}. 
They are the progenitors of the galaxy clusters in the local Universe but are still actively star-forming 
 \citep{Kravtsov2012}, hence are ideal laboratories to study the formation and quenching of massive galaxies. 
 Meanwhile,  these high-$z$ galaxy groups/clusters reside in massive halos and are expected to accrete  cold gas effectively, fueling the star formation in the structures, i.e., the cold accretion theory \citep{Dekel2009, Daddi2021}, as opposed to local galaxy clusters where gas streams are disrupted or heated when entering the hot dark matter halos. Furthermore, the statistics of the number of dark matter halos that host them as a function of redshift and total mass can be used to constrain cosmological parameters and test structure formation theories.  However, observational constraints remain scarce.

Identifying star-forming structures is not straightforward. Radio galaxies  have been used as tracers of proto-clusters \citep[e.g.,][]{Miley2006, Wylezalek2013}, as they tend to reside in massive dark matter halos. However, they do not always trace the structures that host active star formation.  Recently, quite a few Mpc-scale overdensities of emission line galaxies have been discovered \citep[e.g.,][]{Steidel2005, Koyama2013, Toshikawa2018, Lemaux2018}, thanks to the spectroscopy and narrow band imaging of 10m class telescopes. Additionally, massive overdensities are revealed by strong intergalactic Ly$\alpha$ absorption systems in QSO spectra \citep{Cai2017}, and some other tracers are also used \citep[see][]{Overzier2016}
 These extended structures host multiple dark matter halos that may today end up in a single cluster. However, as opposed to single dark matter halos, they cannot be used to  test physical models or theories directly. 
 In this work, we focus on infrared luminous galaxy groups/clusters hosting massive halos at cosmic noon and earlier. 
 
 Submillimeter observations have achieved success in searching for these structures \citep[e.g.,][]{Daddi2009, Capak2011, Walter2012, Casey16, Wang2016, Oteo2018, Miller2018, Zhou2020}, but the samples remain limited and heterogeneous. Besides,  selections via overdensities of submillimeter galaxies alone suffer from projection effects \citep{Chen2023}. Efficient search requires additional constraints on redshift. 
This motivates the  Noema formIng Cluster survEy (NICE), which systematically searches for the massive halos in formation at 2\,$\lesssim$\,$z$\,$\lesssim$\,4. The NICE survey selects overdensities of infrared luminous and massive galaxies based on  deep far-infrared (FIR) and mid-infrared (MIR) priors from \textit{Herschel}/SPIRE and  \textit{Spitzer}/IRAC observations,  
and then spectroscopically confirms them with the gas emission lines detected by NOEMA. 

This paper presents an overview of the NICE survey. Among the first structures confirmed in the NICE survey, this paper also highlights the most distant one, LH-SBC3, which is a starbursting galaxy group with a radio-luminous core at $z$\,=\,3.95.  
Throughout this paper, we adopt a spatially flat $\Lambda$CDM cosmological model with $H_0$\,=\,70\,kms$^{-1}$\,Mpc$^{-1}$, $\Omega_{\rm m}$\,=\,0.3 and $\Omega_{\rm \Lambda}$\,=\,0.7. We assume a \citet{Salpeter1955} initial mass function (IMF) with a conversion factor of 1.7 from the \citet{Chabrier2003} IMF when necessary. All magnitudes are quoted in the AB system \citep{Oke1983}.

\section{Noema formIng Cluster survEy (NICE)}
\label{sec:nice}
\subsection{Survey overview}
The NICE survey is a 159-hour NOEMA large program (PIs: E.\,Daddi and T.\,Wang, proposal ID:M21AA), targeting 48 galaxy overdensities selected from five fields, namely, 
Lockman Hole, Elais-N1, Bo\"{o}tes, XMM-LSS and COSMOS.  25 candidates in the southern sky are complemented by the 40-hour ALMA Cycle8 observations (PI: E.Daddi, project ID:\,2021.1.00815.S) in the ECDFS , COSMOS, and XMM-LSS fields, overlapping with four candidates in the NOEMA observations.
 
\subsection{NICE candidates selection criteria}
\label{sec:sample}
The NICE survey targets starbursting massive galaxy groups/clusters at 2\,$\lesssim$\,$z$\,$\lesssim$\,4. We select overdensities of high-$z$ massive galaxies traced by red IRAC detections  in association with SPIRE-350$\mu$m peakers which show intense star formation. In the following we concisely describe the selection criteria, a detailed introduction will be presented in Zhou et al. (in prep). 

\subsubsection{IRAC selected overdensities}
We use an IRAC color cut  as shown in eq~(\ref{eq:irac}) to select massive galaxies at $z$\,>\,2. This selection criterion is based on the 1.6\,$\mu$m bump in the SEDs of galaxies resulting from  a minimum in the opacity of the H$^-$ ion in the atmospheres of cool stars \citep{John1988}. This method has  proved being efficient in isolating high-$z$ galaxies \citep{Papovich2008, Galametz2012}, and we choose a redder color to select galaxies at higher redshifts. Meanwhile, we constrain the [4.5] magnitude to select massive galaxies while excluding the contamination from stars. In Fig.~\ref{fig:nice-selection}, we show the distribution of the galaxies compiled in the COSMOS2020 catalogue \citep{Weaver2022} and find that most of the massive star-forming galaxies (SFGs) meet our selection criterion. 
\begin{equation}
\begin{split}
\rm [3.6]-[4.5]\,>\,0.1\,; \\
\rm 20<[4.5]\,<\,23\,. \\
\end{split}
\label{eq:irac}
\end{equation}
We used the Nth closest neighbour, $\Sigma_{\rm N}$\,=\,$\frac{N}{\pi\,d_N^2}$, where $d_N$ is the distance to the Nth closest galaxy, as the density estimator. For each field, we created a grid with cell sizes of 3$\arcsec$  and calculate $\Sigma_{\rm N}$ at each cell to construct the surface density map of the color-selected sources.  We then compute the mean and standard deviation of the surface density
across each field. Regions above 5$\sigma$ are  defined as overdensities.   We used both $\Sigma_5$ and $\Sigma_{10}$ and chose targets that satisfied either of those. Fig.~\ref{fig:lh-sbc3-sigma10} (left) shows an example of the $\Sigma_{10}$ map centred on LH-SBC3, which is so far the most distant galaxy group identified in the NICE survey.

\subsubsection{Herschel selected 350$\mu$m peakers}
\label{sec:herschelselect}
\begin{figure*}
\includegraphics[align=c,width=0.34\linewidth]{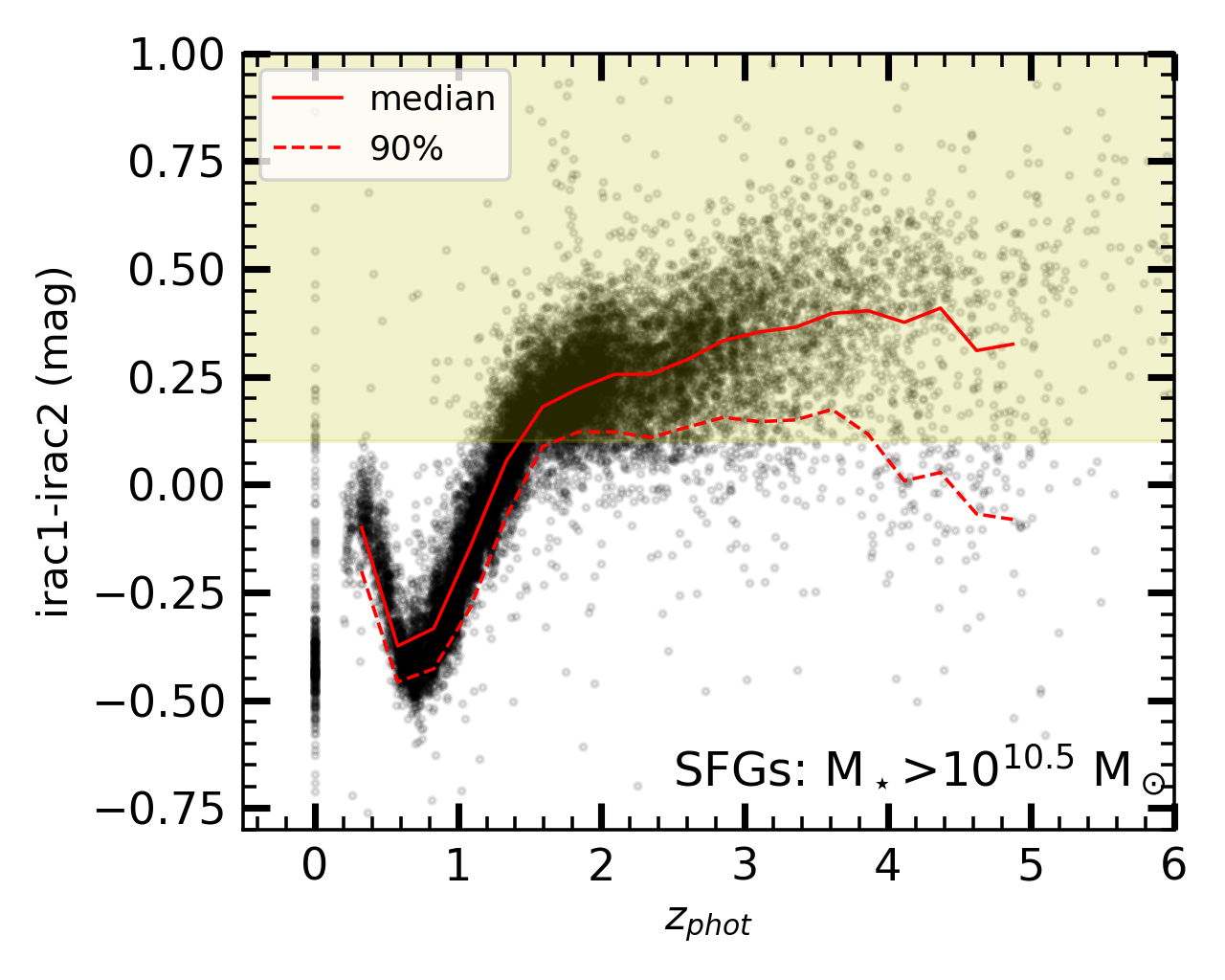}
\includegraphics[align=c,width=0.34\linewidth]{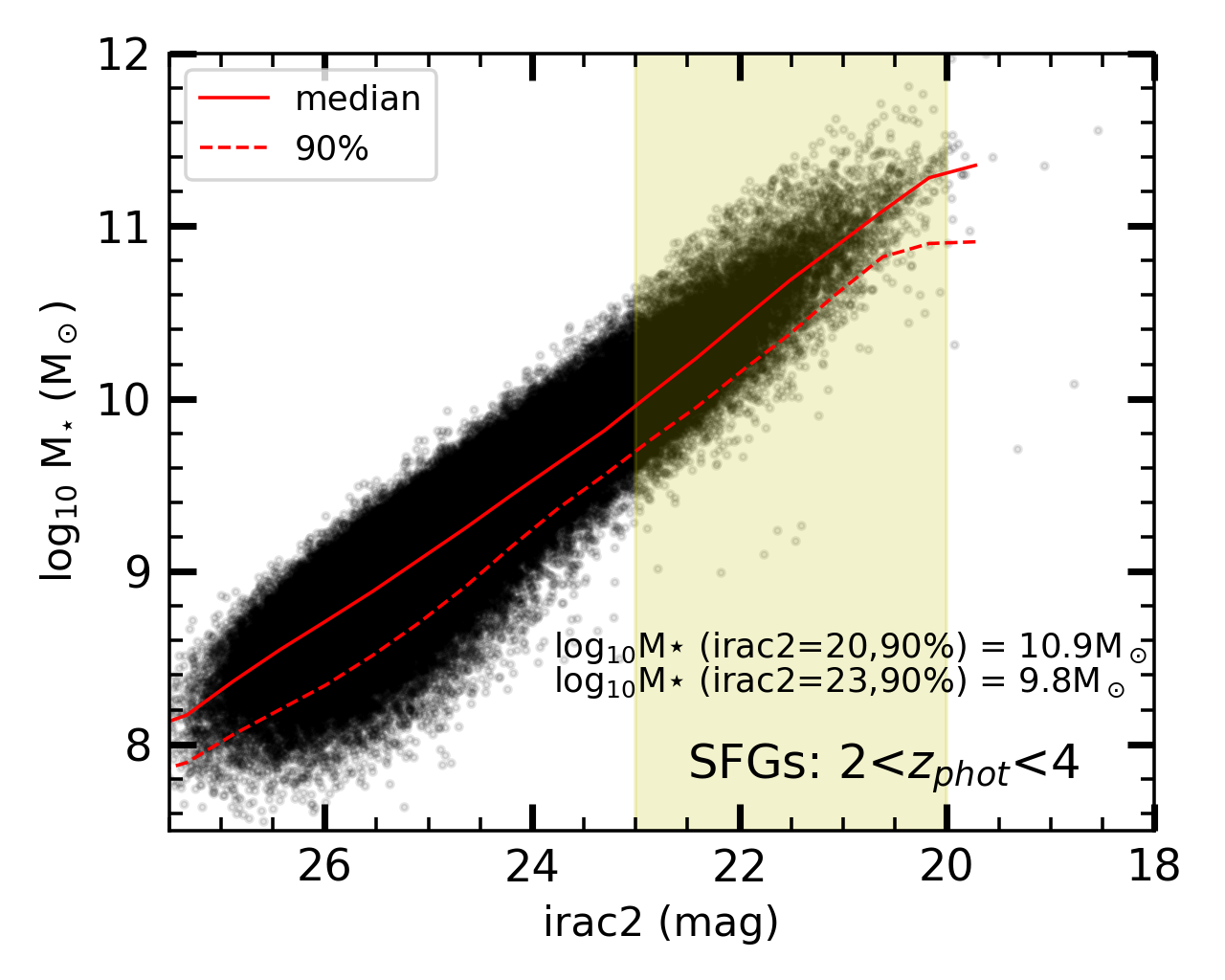}
\includegraphics[align=c,width=0.31\linewidth]{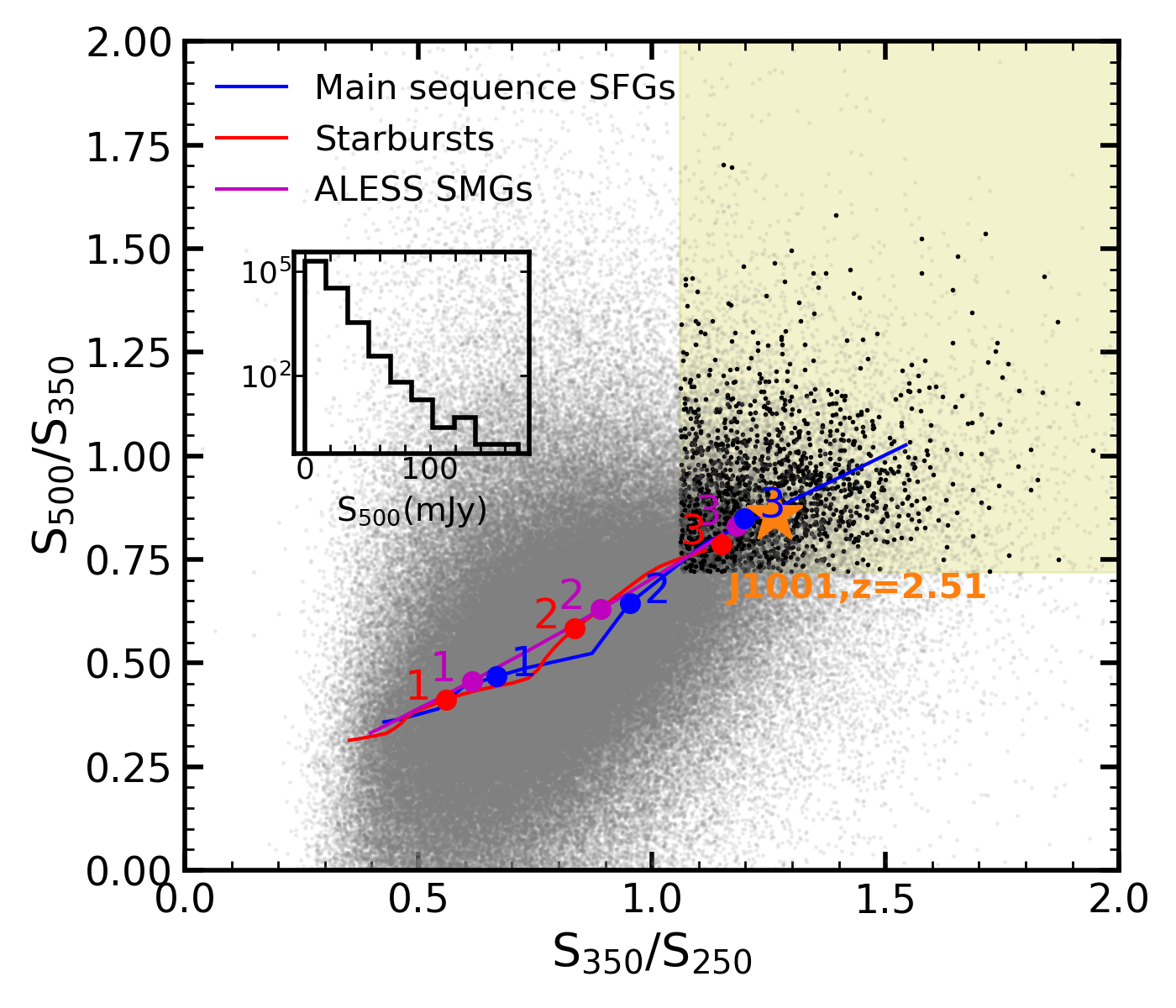}
\caption{NICE candidates selection criteria. \textit{Left}: IRAC color to constrain the redshift to be 2\,$\lesssim$\,$z$\,$\lesssim$\,4. Data points are $UVJ$ color selected massive SFGs from COSMOS2020 \citep{Weaver2022}. Solid and dashed red curves show the median and the lower limit of 90\% of the galaxies. The yellow shade shows the selection criterion. \textit{Middle}: IRAC2 magnitude to choose galaxies more massive than 10$^{10}$\,M$_\odot$. Data points are $UVJ$ color selected SFGs at 2\,<\,$z_{\rm phot}$\,<\,4 from COSMOS2020. The red curves and the yellow shade are the same as in the left panel. \textit{Right}: \textit{Herschel}/SPIRE colors to select starbursting structures at 2\,$\lesssim$\,$z$\,$\lesssim$\,4. Grey dots are  the Herschel detections in all the NICE survey fields, while the black dots meet the selection criteria in eq~(\ref{eq:spire}). The blue, red and magenta curves show the trend of main sequence galaxies, starbursts and SMGs  and the circles are marked by the redshifts accordingly. The starbursting proto-cluster at $z$\,=\,2.51, CL-J1001, is presented as an orange star. See details in Section~\ref{sec:herschelselect}. The inserted histogram displays the distribution of 500$\mu$m fluxes.}
\label{fig:nice-selection}
\end{figure*}

The \textit{Herschel}/SPIRE observations typically have beam sizes of 20$\arcsec$-36$\arcsec$, consistent with spatial scales of massive halos at 2\,$\lesssim$\,$z$\,$\lesssim$\,4.  Therefore, in addition to the IRAC color selection, we verify the collective star formation activities in massive halos with SPIRE colors. 
 The SPIRE color evolution with redshifts for three different types of galaxy SEDs is shown in Fig.~\ref{fig:nice-selection}-right.
The main-sequence and starburst SED templates are based on stacking \citep{Magdis2012}. The template for ALESS $z$\,$\sim$\,2 submillimeter galaxies (SMGs) is taken from \citet{daCunha2015}.  Due to the different PSF sizes of the SPIRE bands, fluxes at longer wavelengths include more sources. As a result, the observed colors of the galaxy clusters/groups are redder than those predicted by individual galaxies. This is corrected based on the comparison between SPIRE colors of the proto-cluster at $z$\,=\,2.51, CL-J1001 \citep{Wang16} and individual galaxies at similar redshifts, which leads to our final sample selection criterion of 2\,$\lesssim$\,$z$\,$\lesssim$\,4 candidates as in eq~(\ref{eq:spire}). 
 We also require sources to be IR bright with $S_{\rm 500}$\,>\,30\,mJy (above 5$\sigma$ confusion noise), which corresponds to $L_{\rm IR}$\,$\sim$\,10$^{13}$\,L$_\odot$ at $z$\,=\,2.5. 
\begin{equation}
\centering
\begin{split}
S_{\rm 500}\,>\,30\,\rm mJy\,; \\
S_{\rm 350}/S_{\rm 250}\,>\,1.06\,; \\
S_{\rm 500}/S_{\rm 350}\,>\,0.72\,. \\
\end{split}
\label{eq:spire}
\end{equation}

\subsection{NOEMA observations}

The NOEMA observations were conducted in band\,1 with two setups. The first setup covers 87.7-95.7\,GHz and 103.2-111.2\,GHz or 80.0-88.0\,GHz and 95.5-103.5\,GHz depending on the estimated redshift of the structures, while the second setup is subsequently arranged regarding the detected lines from the first setup to optimise detections. In general, the observations allow detections of CO(3-2) if they fall at $z$\,$\sim$\,1.9-3.4, or CO(4-3) if they are at $z$\,$\sim$\,2.9-4.8.
The spectra reach a typical rms sensitivity of $\sim$\,130\,$\mu$Jy/beam, and the continuum  reaches $\sim$\,14.5\,$\mu$Jy/beam. Configurations C and D  were adopted to spatially separate the different member galaxies while conserving their total fluxes and then the spatial resolution is $\theta$\,$\sim$\,2$\arcsec$. The starbursting massive galaxy groups/clusters selected by the NICE survey are well covered by  the half-power beam width (HPBW\,$\sim$\,50$\arcsec$).


\begin{figure*}
\centering
\includegraphics[align=c,width=0.47\linewidth]{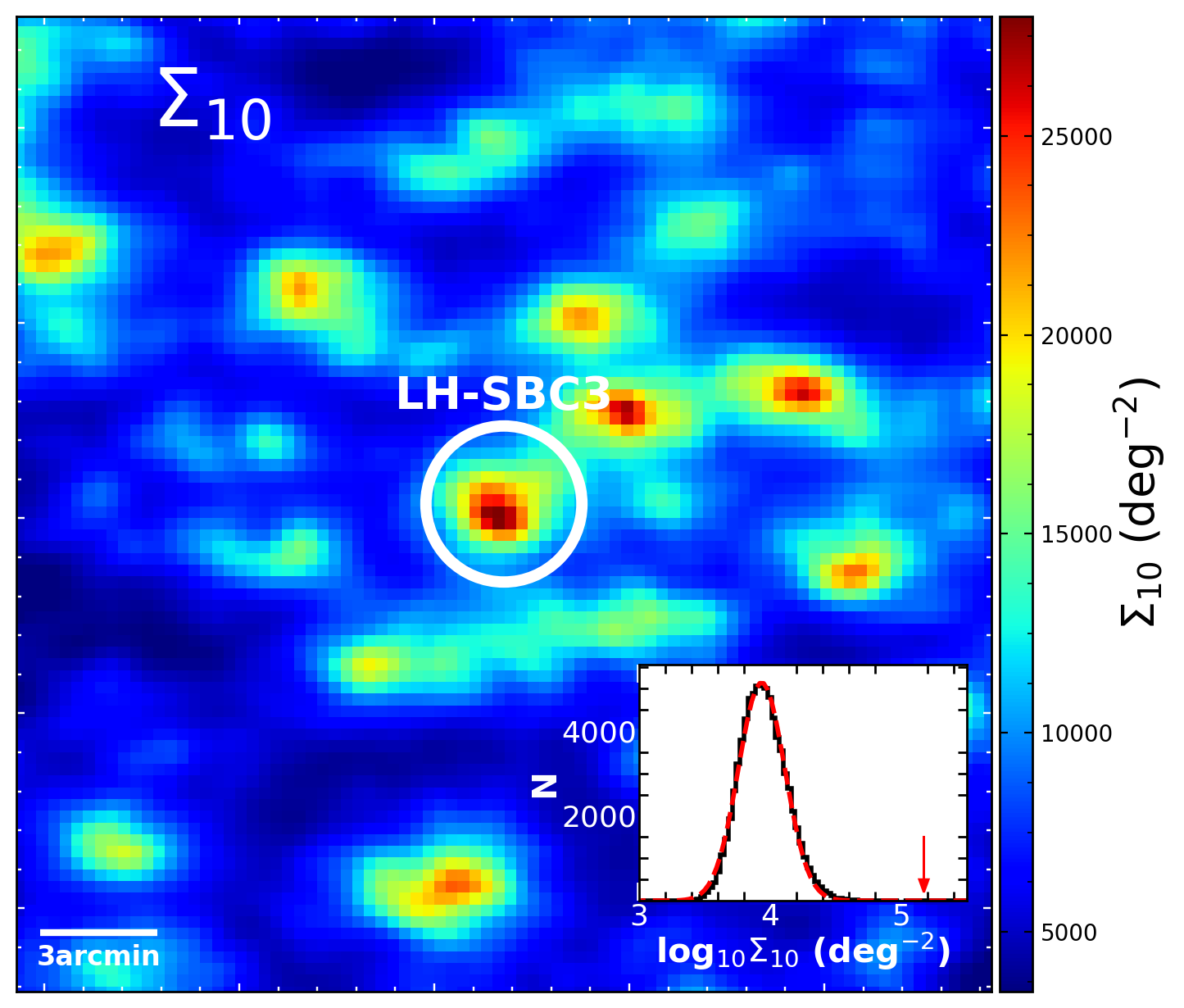}
\includegraphics[align=c,width=0.40\linewidth]{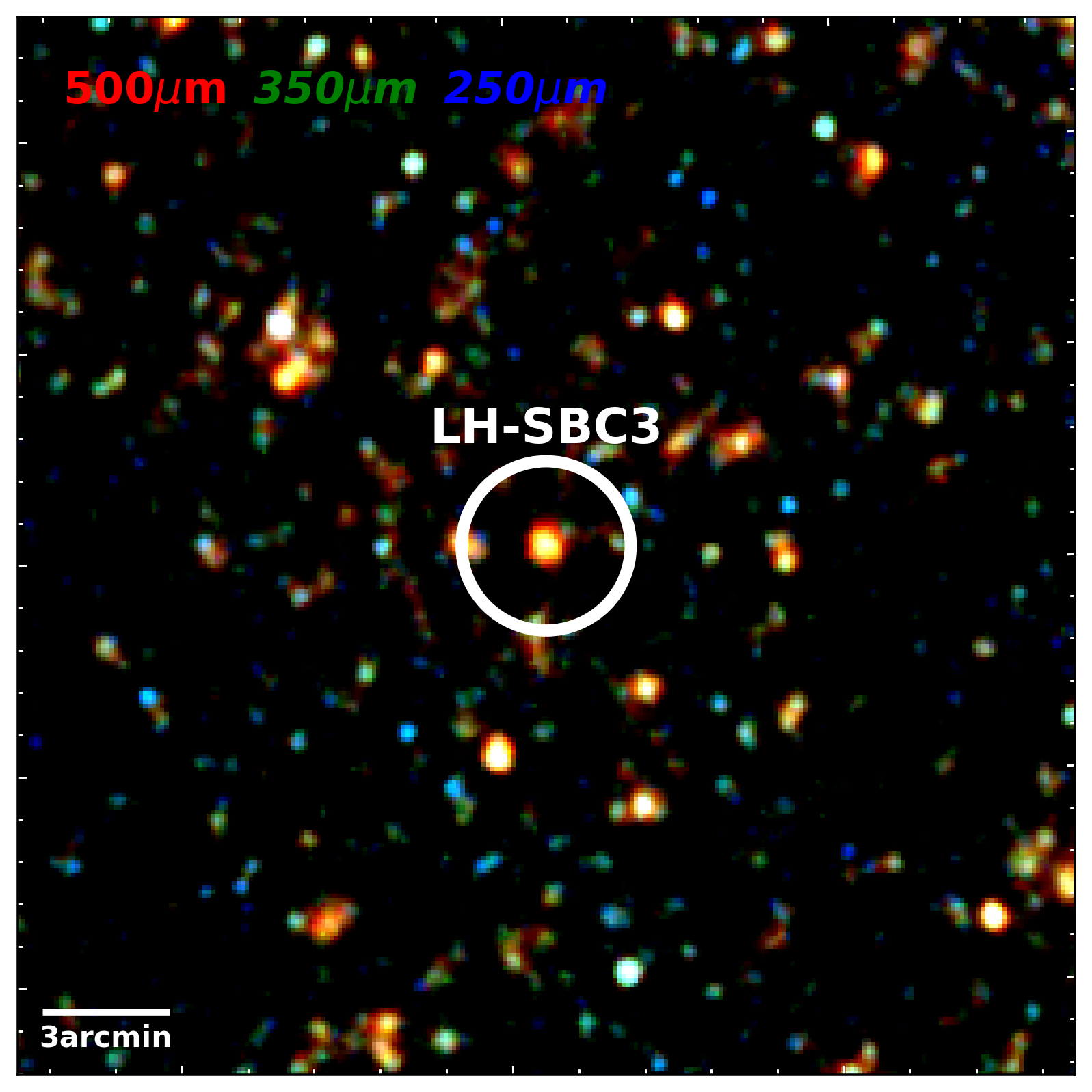}
\caption{LH-SBC3 selected following the method in Section~\ref{sec:sample}. \textit{Left:} Smoothed $\Sigma_{10}$ map of the red IRAC priors (Eq.~\ref{eq:irac}) over the 25$\arcmin$\,$\times$\,25$\arcmin$ area around LH-SBC3. The construction of the map is described in Section~\ref{sec:iracselection}. The inserted histogram is the distribution of $\Sigma_{10}$ (black histogram)  and the Gaussian fit (red dashed curve). The red arrow indicates the peak $\Sigma_{10}$  value of LH-SBC3. \textit{Right:} Herschel/\textit{SPIRE} composite color image of the same area. The R, G, and B channels correspond to  500$\mu$m, 350$\mu$m, and 250$\mu$m, respectively. The white circles in both figures have a radius of 2$\arcmin$. }
\label{fig:lh-sbc3-sigma10}
\end{figure*}



\begin{table*}[ht]
\tiny
\begin{center}
\caption{Line and continuum detections of LH-SBC3 members.}
\label{table:members-line}
\begin{tabular}{lrrrrrrrrrrrrrrrr}
\hline
\hline
\noalign{\vskip 2pt}
  ID & RA &  Dec & $z_{\rm spec}$& $\delta v^\dag$ &  \multicolumn{4}{c}{CO(4-3)}&&\multicolumn{3}{c}{[CI](1-0)}& &  3mm  \\
  \cline{6-10} \cline{12-13} \cline{15-15}\\
\noalign{\vskip -7pt}  
 & & & & &  $I$ & $S/N$  &$P^\ddag$ &$P^\ast_{2000}$  & FWZI$^{\ast\ast}$ && $I$ & $S/N$  && $S$ \\
 & (deg) & (deg) &&(km/s) & (mJy\,km/s) & &   & & (km/s) && (mJy\,km/s) &        && ($\mu$Jy) \\
\hline
\noalign{\vskip 3pt}
$a$ &160.4684 & 58.7558 & 3.949 & -54& 1567$\,\pm\,$57 &27.5&0.000&0.000&825& &205$\,\pm\,$68 &3.0   && 73$\,\pm\,$11 \\
\noalign{\vskip 2pt}
$b$ &160.4764 & 58.7584 & 3.954 & -348& 528$\,\pm\,$45 &11.6 &0.000&0.000&671& &136$\,\pm\,$54 &2.5   && 114$\,\pm\,$10 \\ 
\noalign{\vskip 2pt}
$c$ & 160.4648 & 58.7554 & 3.940 &  543& 180$\,\pm\,$38 &4.8&0.024 &0.001&566& &<108 &<2   && 23$\,\pm\,$10 \\
\noalign{\vskip 2pt}
$d$ & 160.4702 & 58.7561   & 3.951 &-141& 226$\,\pm\,$52 &4.6&0.034&0.002&824& &190$\,\pm\,$65 &2.9  & &43$\,\pm\,$10 \\
\noalign{\vskip 1pt}
\hline 
\end{tabular}
\end{center}
    \begin{tablenotes}
      \tiny
      \item  \textbf{Notes.}  $\delta v^\dag$: velocity relative to the cluster center by assuming the average redshift $z_{av}$\,=\,3.949; $P^\ddag$: chance of spurious probability calculated over the entire spectrum  based on \citet{Jin2019}; $P^\ast_{2000}$: chance of spurious probability calculated over $\pm$2000\,km/s of the CO(4-3) line at $z_{av}$\,=\,3.949;   FWZI$^{\ast\ast}$: full width at zero intensity of the lines. 
    \end{tablenotes}
\end{table*}


\begin{table}[ht]
\begin{center}
\caption{Physical properties of  LH-SBC3 members.}
\label{table:members}
\begin{tabular}{lrrrrrrrrrrrrrrr}
\hline
\hline
\noalign{\vskip 3pt}
  ID &  $S_{\rm150MHz}$ &  log$_{10}$$M_\star$ & SFR \\
\noalign{\vskip 2pt}
     &  ($\mu$Jy) &  (M$_\odot$) & (M$_\odot$/yr) \\
\hline
\noalign{\vskip 3pt}
$a$   & 1519$\,\pm\,$46 & 11.10$^{+0.10}_{-0.17}$ & 1247$\,\pm\,$82\\
\noalign{\vskip 2pt}
$b$ & 68$\,\pm\,$33    & 10.58$^{+0.11}_{-0.15}$  & 1947$\,\pm\,$128 \\
\noalign{\vskip 2pt}
$c$ & 59$\,\pm\,$33    & 10.75$^{+0.37}_{-0.24}$  & 392$\,\pm\,$26 \\
\noalign{\vskip 2pt}
$d$    & <66           & 10.23$^{+0.66}_{-1.63}$  & 734$\,\pm\,$48  \\
\noalign{\vskip 3pt}
\hline 
\end{tabular}
\end{center}
    \begin{tablenotes}
      \small
      \item  \textbf{Notes.} Fluxes at 150MHz are measured by LeBail et al. (in prep) using \textsc{pybdsf}\footnote{\url{https://pybdsf.readthedocs.io/en/latest/}}. Stellar masses are derived from optical to mid-IR SED fitting assuming a delayed star formation history using \textsc{fast++}. SFRs are obtained by scaling the total SFR of the structure with the 3mm fluxes of individual members. See details in Section~\ref{sec:properties}.  
    \end{tablenotes}
\end{table}

\subsection{NOEMA data reduction and source extraction}
The NOEMA data were calibrated and reduced using \texttt{GILDAS}\footnote{\url{ https://www.iram.fr/IRAMFR/GILDAS}}. We produced $uv$ tables following the standard NOEMA data reduction with \texttt{GILDAS CLIC} package\footnote{\url{https://www.iram.fr/IRAMFR/GILDAS/doc/pdf/pdbi-cookbook.pdf}}, which includes bandpass, phase, amplitude and flux scale calibration, and data flagging. We then made dirty and clean images using \texttt{GILDAS MAPPING} \texttt{go uvmap} and \texttt{go clean} tasks; meanwhile, we extracted spectra for multiple sources simultaneously in the $uv$ plane using the \texttt{run uv$\_$fit} task. This \texttt{GILDAS uv$\_$fit} task can model analytical source models in the image plane and convert the emission to the Fourier space then directly fit the observed visibilities. This is similar to the \texttt{CASA} \texttt{uvmodelfit}, but the latter cannot fit multiple sources simultaneously.  The phase centres were set as the coordinates of the NICE candidates, namely, the coordinates of the corresponding SPIRE-350$\mu$m peakers. Primary beam attenuation correction was estimated for each source at their corresponding positions in the image, based on the \texttt{GILDAS MAPPING go primary} task  with the correction factors ranging from 1.01 to 1.22, and the corrections were applied to the line and continuum measurements. Natural weighting was adopted in the imaging procedures to maximize the sensitivity and a cleaning threshold of 2.5$\sigma$ is used when running the \textsc{gildas mapping} \texttt{go clean} task.
We first extracted the spectra of continuum detections. 
If a source is spatially resolved, we use a Gaussian model to fit the uvdata and then extract the spectra.  The unresolved sources were extracted by fitting point sources. The same procedure is also performed at the IRAC position of each cluster member candidate. 
After combining the spectra from two observation setups, we run the line-searching algorithm as in \citet{Coogan2019}  over each 1D spectrum. We first measured the line-free continuum using a noise-weighted fit, assuming a slope of 3.7 in frequency and obtain the continuum flux. We then subtracted the continuum from the spectra and use \textsc{mpfit}\footnote{\url{http://cow.physics.wisc.edu/~craigm/idl/idl.html}} to fit the lines with Gaussians. We obtained the redshift from the first line considering the photometric of the galaxies and then fixed the redshift and measured the second.

\section{LH-SBC3: A galaxy group at $z$\,=\,3.95}
\label{sec:lh-sbc3}
LH-SBC3 is a galaxy group at $z$\,=\,3.95 in the Lockman Hole field. 
Among the first NICE structures that are spectroscopically confirmed, it is also the most distant one. We present it here as a highlight.

\begin{figure*}
\centering
\includegraphics[align=c,width=0.68\linewidth]{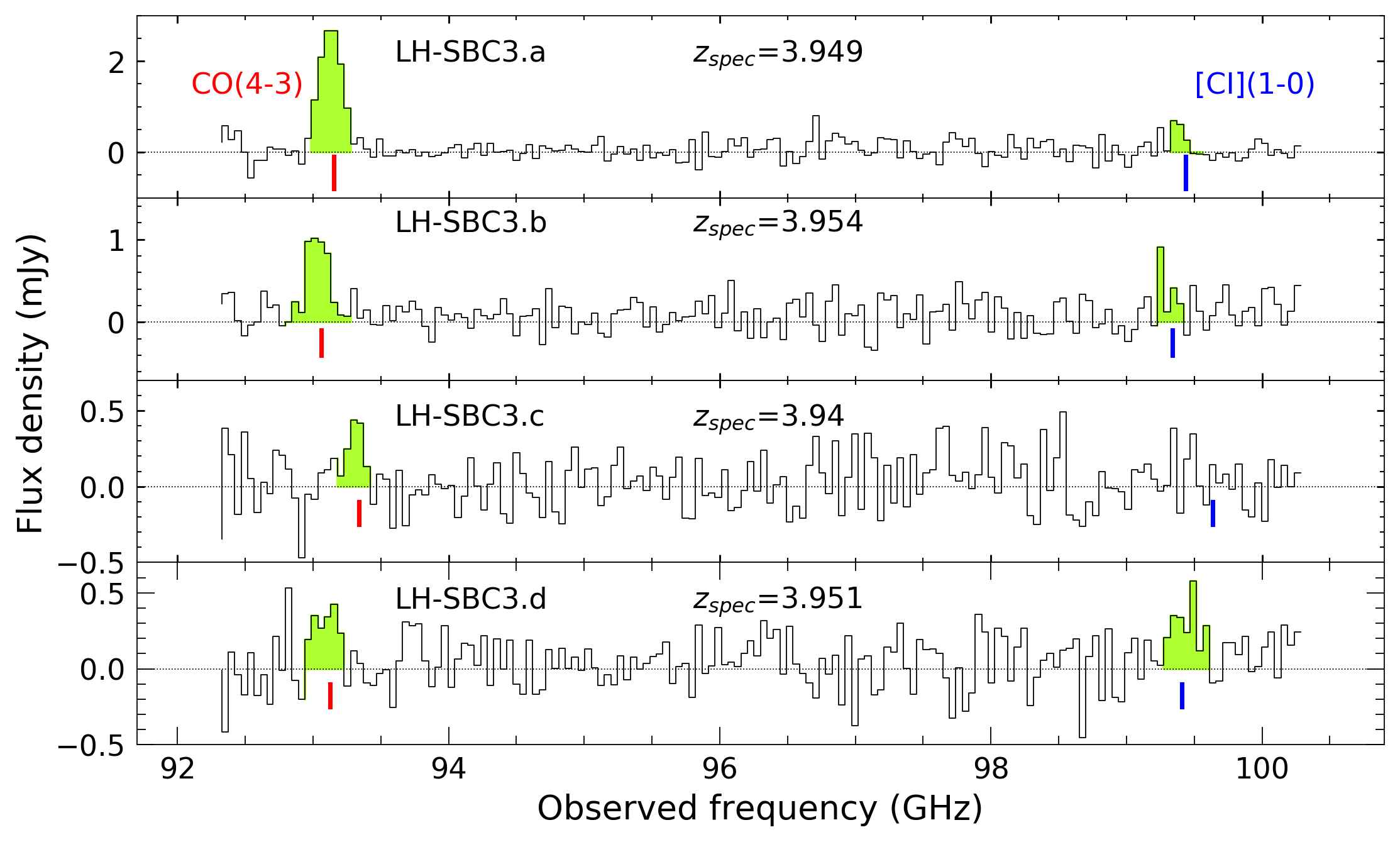}
\includegraphics[align=c,width=0.29\linewidth]{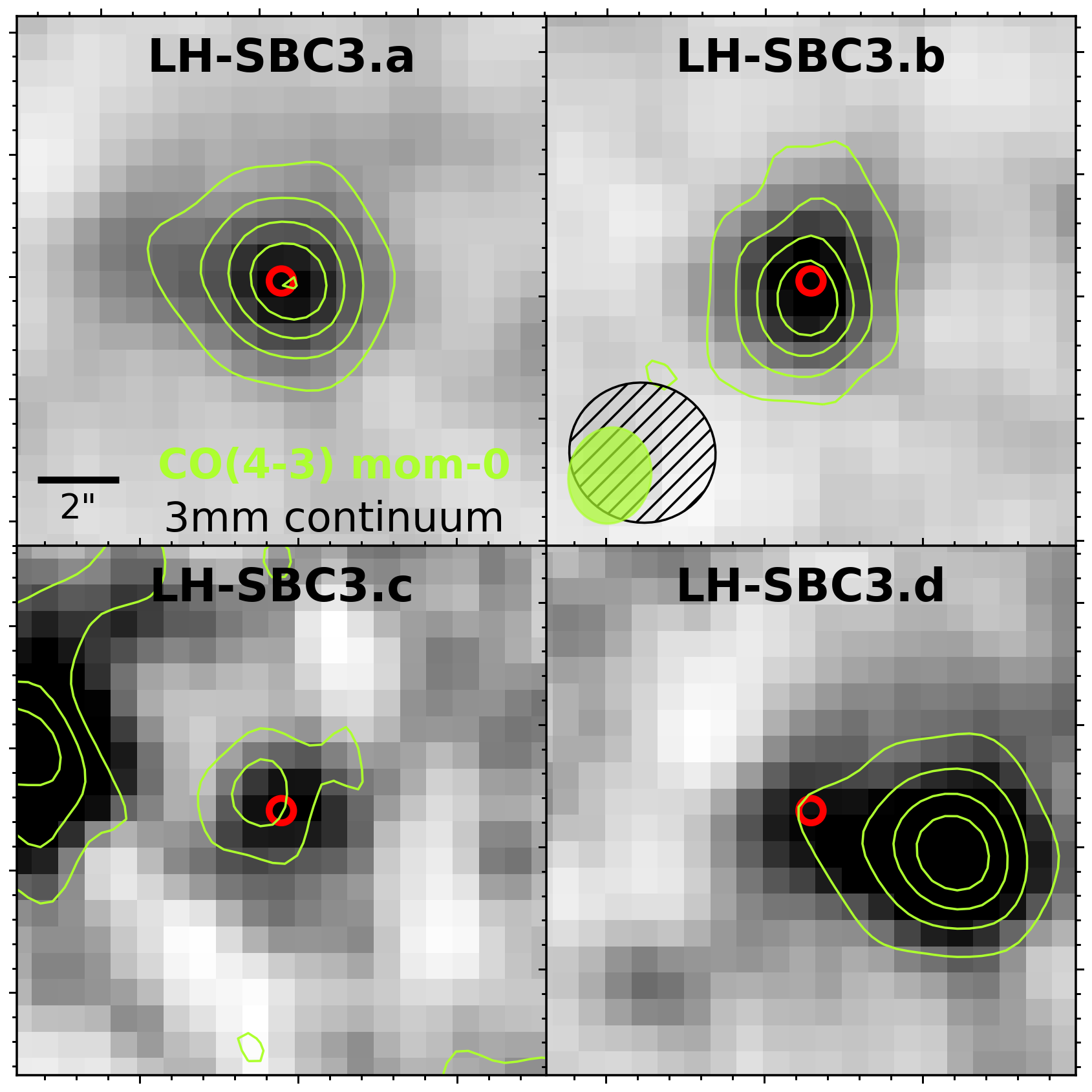}
\caption{The four members of LH-SBC3 spectroscopically confirmed by NOEMA. \textit{Left}: NOEMA spectra of four confirmed cluster members. The channel width of the spectra is rebinned to be $\sim$150\,km\,s$^{-1}$.  The detected lines are highlighted in yellow. CO(4-3) and [CI](1-0) frequencies are marked by vertical bars in red and blue, respectively. \textit{Right}: 3mm continuum images of the four members overlaid by CO(4-3) intensity contours in yellow. The red circles denote the positions of the galaxies determined by the flux peak of the 3mm continuum. The CO(4-3) intensity contours start at 1$\sigma$ (equivalent to $\sim$56 mJy\,km$^{-1}$\,beam$^{-1}$) in  steps of 1$\sigma$. Beams of the 3mm and CO(4-3) intensity maps are shown in the second panel as black-hatched and yellow patterns corresponding to 3.43\arcsec$\times$3.60\arcsec and 2.37\arcsec$\times$2.02\arcsec, respectively. 
} 
\label{fig:sbc3}
\end{figure*}

LH-SBC3 consists of four members detected with CO(4-3), LH-SBC3.a, b, c, and d (as shown in Fig.~\ref{fig:sbc3}).  The two brightest ones, LH-SBC3.a and  b, also show [CI](1-0) with which we robustly confirm their redshifts. The line and continuum detections are listed in Table~\ref{table:members-line}. 
LH-SBC3.a and d are two spatially resolved  peaks in the continuum at 3mm. They are around 3$\arcsec$ apart, larger than the typical size of a massive star-forming galaxy at $z\sim$\,4.  We will also show in Section~\ref{sec:radio} that LH-SBC3.d has marginal detection at 150MHz 3$\arcsec$ from LH-SBC3.a. 
Two more sources have emission lines detected at $f_{\rm obs}$\,$\sim$108\,GHz.  Based on their photometric redshifts, they are most likely at $z$\,$\sim$\,2.2 and the lines correspond to  CO(3-2). They could also be at $z$\,$\sim$\,3.2 if the lines are CO(4-3). In either case, they may be part of a foreground structure. The spectra of these two galaxies are shown in the appendix (Fig.~\ref{fig:foreground}).

\begin{figure*}
\centering
\includegraphics[align=c,width=0.75\linewidth]{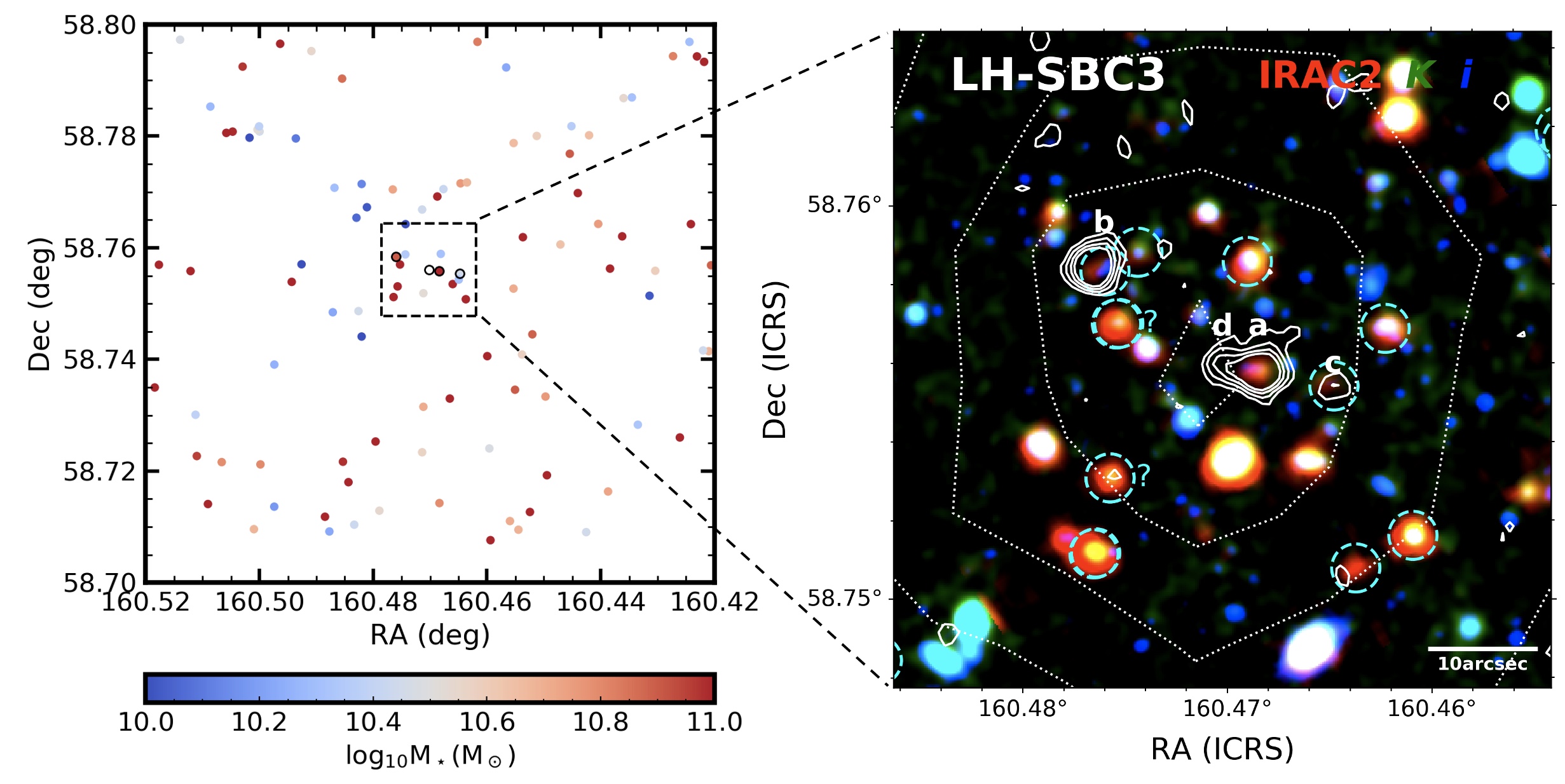} 
\caption{
\textit{Left:}  Spatial distribution of galaxies with photometric redshift intervals consistent with $z$=3.95. The dashed box denotes the  1$\arcmin$\,$\times$\,1$\arcmin$($\sim$\,2\,cMpc\,$\times$\,2\,cMpc) region of the galaxy overdensity, LH-SBC3. The dots are color-coded by their stellar masses. 
\textit{Right:}  Zoom-in RGB image (red: IRAC2; green: \textit{K}; blue: \textit{i}) of LH-SBC3. The white solid contours denote the 3mm continuum, starting at 2$\sigma$ in steps of 1$\sigma$. The four CO-confirmed members are  labelled with their IDs. Note: LH-SBC3.a and d are very close. The cyan dashed circles are the IRAC-color selected priors. The priors labelled with a question mark could come from another structure at $z$\,=\,3.25 or $z$\,=\,2.19, see discussion in Section~\ref{sec:lh-sbc3}. The dotted white contours show 500\,$\mu$m flux at 20, 40, 60 mJy/beam levels. 
} 
\label{fig:sbc3-rgb}
\end{figure*}

LH-SBC3 was selected to have a 7$\sigma$ excess in $\Sigma_{10}$ when searching in the entire Lockman Hole field  (Fig.~\ref{fig:lh-sbc3-sigma10}-\textit{Left}).  It contains six IRAC priors as selected by Eq.~(\ref{eq:irac}) within a radius of  25$\arcsec$ ($r$\,$\sim$\,180\,pkpc, Fig.~\ref{fig:sbc3-rgb}). 
The four NOEMA detections were not selected as priors because they are intrinsically faint (high dust obscuration) at the IRAC wavelength and tend to be blended in this crowded region, which increases the uncertainty of their photometry, and IRAC colors. 
We supplement the existing images with additional $H$, $K$ observations from CFHT (PI: T. Wang) and use the deeper NIR images for source extraction.
 This is to avoid the dilution of these optically faint IRAC priors in the detection image. 
Finally, we confirmed that LH-SBC3.b and LH-SBC3.c should have been selected while LH-SBC3.a and LH-SBC3.d are too close to de-blend and hence missed. 
Besides, the spatial distribution of galaxies with photometric redshift intervals consistent with $z$\,$\sim$\,3.95  shows an apparent excess of massive galaxies centred on LH-SBC3 (Fig.~\ref{fig:sbc3-rgb}), which reinforces the existence of an overdensed structure here.

\begin{figure}
\centering
\includegraphics[align=c,width=1.1\linewidth]{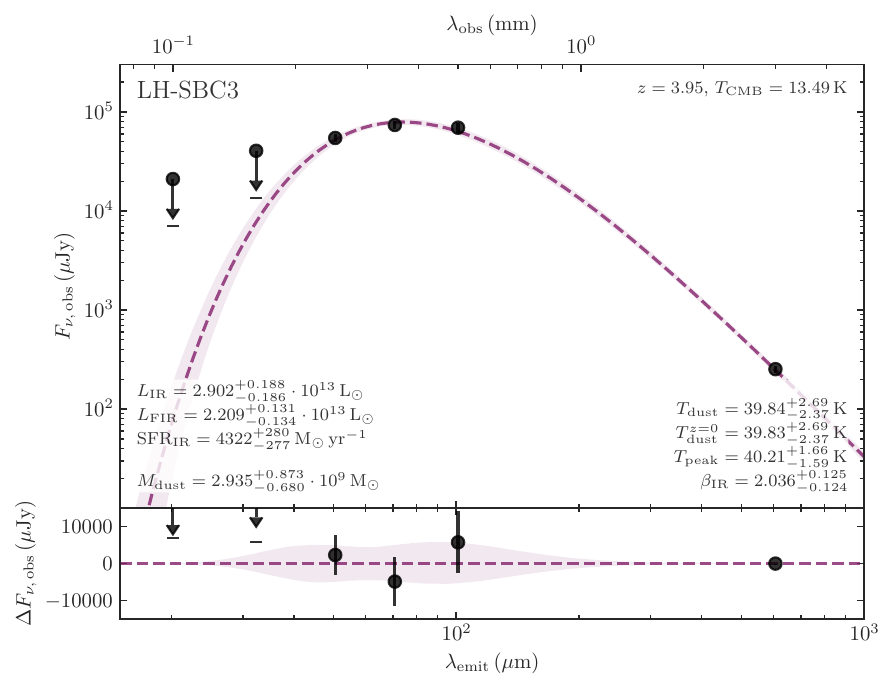}
\caption{FIR SED of LH-SBC3. Line-of-sight contamination to the SPIRE fluxes is subtracted using the deblending procedures in \citet{Jin2018}. The 3mm flux is well  fitted by an optically thin dust model \citep{Witstok2022}. } 
\label{fig:sbc3-firsed}
\end{figure}

LH-SBC3 shows strong emission in the FIR (Fig.~\ref{fig:lh-sbc3-sigma10}, right panel) 
indicating intensive star formation. After deblending the foreground sources from the SPIRE images following \citet[][also see \citealt{Liu2018}]{Jin2018}, we obtained the SPIRE fluxes of the structure to be $S_{\rm 250}$\,=\,54.6$\pm$5.3\,mJy, $S_{\rm 350}$\,=\,73.8$\pm$6.7\,mJy, and $S_{\rm 500}$\,=\,69.3$\pm$8.4\,mJy. In Fig.~\ref{fig:sbc3-firsed}, we find that the total flux of the four members at 3mm is well fitted by an optically thin dust model with MERCURIUS\footnote{\url{https://github.com/joriswitstok/mercurius }} as in \citet{Witstok2022}.  Therefore we rule out the line-of-sight interlopers contributing to the \textit{Herschel}/SPIRE fluxes.




\begin{figure}
\centering
\includegraphics[width=\linewidth]{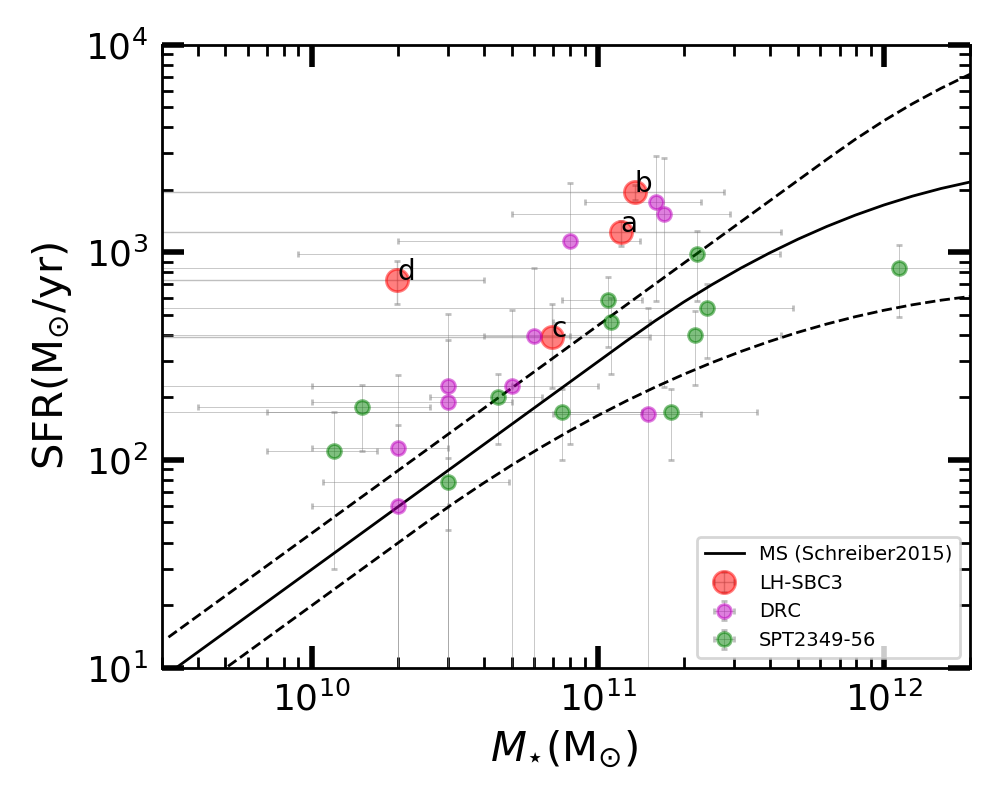}
\caption{Position of the LH-SBC3 member galaxies (red circles) with respect to the star-forming main sequence at $z$\,=\,4 \citep{Schreiber2015}. Galaxies in the Dense Red Core \citep[DRC, $z$=4][]{Oteo2018, Long2020} and SPT2349-56 \citep[$z$\,=\,4.3,][]{Miller2018, Hill2022} are shown as magenta and green circles, respectively.}
\label{fig:sbc3-physics}
\end{figure}

\subsection{Properties of LH-SBC3}
\label{sec:properties}
We fit the optical to mid-infrared SEDs using \textsc{fast++}\footnote{\url{https://github.com/cschreib/fastpp}}, \textsc{cigale} \citep{Boquien19} and \textsc{bagpipes} 
 \citep{Carnall2018} to derive the stellar masses ($M_{\star}$). All codes give consistent results. We assume delayed star formation history and a \citet{Calzetti00} attenuation curve in the fitting. If assuming constant star formation history, the derived masses would be no more than 0.3\,dex lower.
Three of the four members are massive, with $M_\star$>10$^{10.5}$\,M$_\odot$ (Table~\ref{table:members}).
We calculate SFR by 1) fitting the mid-IR to 3mm fluxes with \textsc{cigale}. For LH-SBC3.a and LH-SBC3.b, we include the deblended SPIRE fluxes following the procedure described in \citet{Jin2018}. 2) scaling the 3mm fluxes of the individual members with the total flux of the entire structure that can be used to derive the total SFR of the group since we have confirmed in Fig~\ref{fig:sbc3-firsed} that the deblended SPIRE fluxes and 3mm flux of LH-SBC3 match the dust emission model well. 3) Using the CO(4-3) -- $L_{\rm IR}$ scaling relation \citep{Liu2015}, which considers dense molecular gas emitting at high-$J$ as good tracers of star formation, we estimated the SFRs again. The three methods give generally consistent results. Hereafter we use the scaled SFRs as in method 2.

The four CO detected members are concentrated in an area of $r$\,<\,0.2\,pMpc, and show velocity dispersion within 500\,km/s of the average redshift, then LH-SBC3 is consistent with being a single massive halo.
We estimate the halo masses ($M_{\rm halo}$) of LH-SBC3 with two different methods. 1)  Adopting the $M_{\rm halo}$-$M_{\star}$ scaling relation in \citet{Behroozi13}, the most massive member of LH-SBC3 gives an upper limit of $M_{\rm halo}$\,$\sim$\,$10^{13.2}$\,M$_\odot$. 
2) We first derive the total stellar mass of the galaxies with stellar mass $M_\star$\,>\,$10^{10.5}$\,M$_\odot$ in LH-SBC3 to be $10^{11.37}$\,M$_\odot$  if we count only the CO detected members, or  $10^{11.68}$\,M$_\odot$ if we include all the galaxies with $z_{\rm phot}$\,$\sim$\,3.95 as shown in Fig.~\ref{fig:sbc3-rgb}. We then  extrapolate a total stellar mass
down to $10^7$\,M$_\odot$ assuming the stellar mass function of field galaxies from \citet{Muzzin2013} at
3\,$\le$\,$z$\,<\,4, and obtain $10^{11.41}$ - $10^{11.72}$\,M$_\odot$ . Finally, we adopt the scaling relation between total stellar and halo mass derived from $z$\,$\sim$\,1 clusters with halo masses in the range 0.6-16\,$\times$\,$10^{14}$\,M$_\odot$ \citep{vanderBurg2014}  and  yield a halo mass of $M_{\rm 200}$\,$\sim$\,10$^{12.75}$-10$^{13.28}$\,M$_\odot$. 

We further measure the halo mass-normalized integrated SFR ($\Sigma$SFR/$M_{\rm halo}$) of LH-SBC3. Assuming that the star formation in the structure is dominated by the four NOEMA detected members, we obtain $\Sigma$SFR/$M_{\rm halo}$\,$\sim$\,2-10$\,\times$10$^{-10}\,$yr$^{-1}$, around 2\,dex higher than the clusters at $z$\,=\,1--2 \citep{Alberts2016}. Compared to clusters at lower redshifts,  we find that $\Sigma$SFR/$M_{\rm halo}$ rises rapidly with redshift, as predicted by \citet{Alberts2014} (see more discussion in e.g., \citealt{Popesso2015, Alberts2022}). This indicates a vigorous star formation in the structures in the early Universe.  We 
speculate that LH-SBC3 is still in the growing phase and will evolve into a transitioning phase like CL J1001 at $z$\,$\sim$\,2.5 \citep{Wang16}.

In Fig.~\ref{fig:sbc3-physics}, we compare member galaxies of LH-SBC3 with those in two other  starbursting proto-clusters at $z\sim$\,4 Dense Red Core \citep[DRC,][]{Oteo2018}, and SPT2349-56 \citep{Miller2018}. The LH-SBC3 member galaxies generally fall within the scatter shown by DRC and SPT2349-56. The two strongest CO emitters, LH-SBC3.a and LH-SBC3.b are massive and starbursting.



\subsection{A radio luminous core}
\label{sec:radio}
\begin{figure}
\centering
\includegraphics[width=0.99\linewidth]{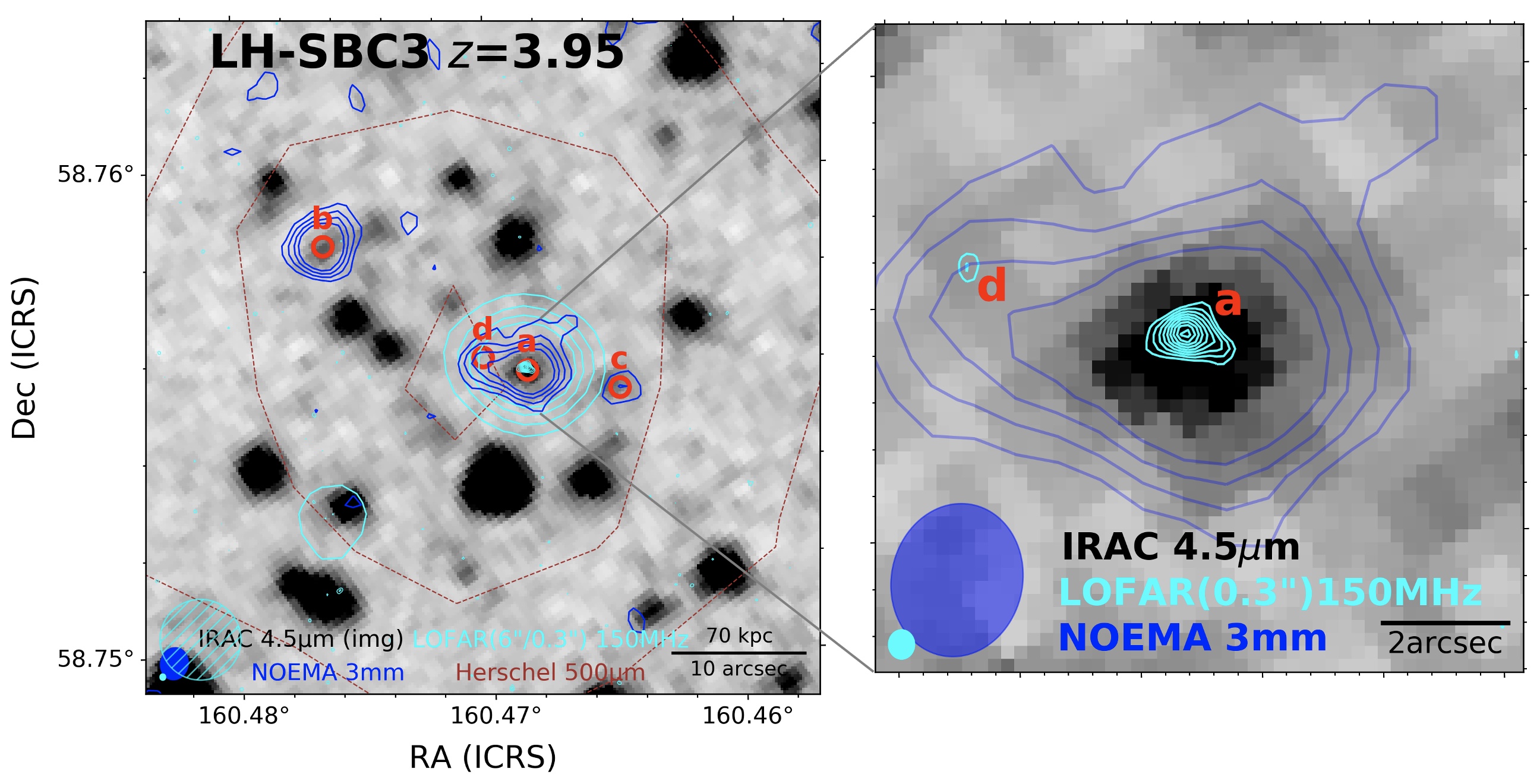}
\caption{A radio-luminous core in LH-SBC3. \textit{Left}:  IRAC2 image of the LH-SBC3 core area overlaid by the contours of NOEMA 3mm continuum (blue),  LOFAR 150MHz (cyan) and SPIRE 500$\mu$m (brown dashed). The four confirmed members are marked as red circles and labelled with their IDs. The 3mm contours start at 2$\sigma$ in steps of 1$\sigma$. The thick cyan contours show the 150MHz continuum at 6$\arcsec$ starting at 3$\sigma$ in  steps of 1$\sigma$. The 500$\mu$m  contours are at 20, 40, 60 mJy/beam levels. Beams of the 3mm map,  the LOFAR 150MHz at  0.3$\arcsec$ and the  LOFAR 150MHz  map at 6$\arcsec$  are shown in the bottom left corner in blue,  cyan, and hatched cyan  patterns, respectively.
\textit{Right}: Zoom-in on the radio galaxy LH-SBC3.a. The cyan contours represent the 150MHz continuum at 0.3$\arcsec$ starting at 3$\sigma$ in  steps of 1$\sigma$. Note that LH-SBC3.d. shows marginal detection of $\sim3\sigma$, which reinforces that it is a separate galaxy from LH-SBC3.a.
}
\label{fig:sbc3-radio}
\end{figure}

The most massive of the four CO-confirmed members, LH-SBC3.a, shows strong radio emission at 150MHz, $S_{\rm 150MHz}$\,=\,1519\,$\mu$Jy. Comparing the fluxes in the high (0.3$\arcsec$) and low (6$\arcsec$) spatial resolution images \citep{Sweijen2022, Tasse2021}, we find that all the radio emission originates from the core of this galaxy (Fig~\ref{fig:sbc3-radio}). This galaxy is also detected at 324.5\,MHz (corresponding to 1.6\,GHz in the rest-frame) by the VLA, with $S_{\rm 324.5MHz}$\,=\,888\,$\mu$Jy \citep{Owen2009}. This gives a radio spectral index of $\alpha$\,=\,$-$0.7 and we derive a radio luminosity of $L_{\rm 1.4GHz, rest}$\,=\,3.0$\times$10$^{25}$\,W\,Hz$^{-1}$.  We find that the radio luminosity is $\sim$5 times higher than expected from star formation \citep{Delhaize17,Delvecchio2021}, indicating the existence of a radio AGN, but it does not show any   signs of other AGN in diagnostics available (optical or MIR or X-ray). This massive galaxy resides at the centre of the structure, and it is likely merging/interacting with LH-SBC3.d, whose CO(4-3) shows an offest of 155\,km/s, therefore it is probably a future BCG in the cluster. Two other members, LH-SBC3.b and LH-SBC3.c also show moderate radio continuum emission ($S_{\rm 150MHz}$\,=\,68 and 59 $\mu$Jy), but they both follow the IR-radio correlation of normal star-forming galaxies.

Powerful radio AGNs are also revealed in DRC and SPT2349-56 \citep{Oteo2018, Chapman2023}. Among them, DRC-6 has a radio luminosity similar to LH-SBC3.a, but it is not at the center of the structure. The radio source found in SPT2349-56 is extremely bright, $L_{\rm 1.4GHz, rest}$\,=\,(2.4$\pm$0.3)$\times$10$^{26}$\,W\,Hz$^{-1}$, 8$\times$ brighter than LH-SBC3.a, but this could result from blending several galaxies in the large beams of ATCA and ASKAP. Moreover, in these two cases, the radio galaxies are the main sequence galaxies, unlike the massive and starbursting LH-SBC3.a. Radio-loud galaxies have been used to trace high-$z$ structures. Compared to the radio loud tracers in the CARLA survey \citep{Wylezalek2013}, we find the rest-frame 500 MHz radio luminosity of LH-SBC3.a is at least 1.7 dex lower.  Therefore, the 
starbursting radio core in LH-SBC3 may indicate an early stage of the radio-loud phase when the negative feedback, such as heating from the AGN has not yet come into effect, or it may not evolve to a more powerful stage to impact the host galaxy or the environment negatively. High-redshift radio galaxies (HzRGs) are found to be the most massive galaxies and probable progenitors of BCGs in galaxy clusters (see \citealt{Miley2008} for a review). LH-SBC3, together with DRC and SPT2349-56, provides evidence of radio galaxies with vigorous star formation in proto-cluster cores, and the NICE survey will  facilitate statistical studies on the probable evolution of radio galaxies into BCGs.

\section{Summary}
In this paper,  we introduce the Noema formIng Cluster survEy (NICE) survey, a 159-hour NOEMA large program to  search for starbursting massive galaxy groups/clusters at $z$\,>\,2. We select candidates by associating SPIRE color-selected IR luminous sources with prominent overdensities of IRAC color-selected massive high-redshift galaxies. This survey will provide us with a statistically large sample to better understand the formation of these structures as well as their influence on the formation and evolution of the galaxies therein.

We verify the feasibility of the NICE candidate selection method by reporting the discovery of a galaxy group at $z_{\rm spec}$=3.95, LH-SBC3 from the first quarter of the survey observations.  It is spectroscopically confirmed by four galaxies with CO(4-3)  and [CI](1-0) line detections within 20$\arcsec$\,$\times$\,20$\arcsec$ (Fig.~\ref{fig:sbc3}). All these galaxies are gas-rich and three of them are massive, $M_\star$\,$>$\,10$^{10.5}$\,M$_\odot$. The most massive one, LH-SBC3.a,  is a prominent starburst and resides at the centre of the structure. It also hosts a radio AGN with $L_{\rm 1.4GHz, rest}$\,=\,3.0$\times$10$^{25}$\,W\,Hz$^{-1}$. 
 The integrated SFR of LH-SBC3 reaches $>$\,4000\,M$_\odot$/yr, with all the CO-detected members being star-forming and two of them lying significantly above the star-forming main sequence. We estimate the halo mass to be $\sim$10$^{13}$\,M$_\odot$, and derive a halo mass-normalized total SFR of $\sim$\,2-10$\,\times$10$^{-10}\,$yr$^{-1}$, which is around 2\,dex higher than the clusters at $z$\,=\,1--2.
 This suggests that the specific SFR per halo mass of clusters continues to grow back in cosmic time,  up to $z$\,$\sim$\,4. Together with a few previously discovered similar structures at  $z$\,$\sim$\,4, this finding indicates that these compact groups or starbursting cluster cores represent an important phase of massive galaxy assembly in cluster formation. 

\begin{acknowledgements}
L.Z. and Y.S. acknowledge the support from the National Key R\&D Program of China No. 2022YFF0503401, the National Natural Science Foundation of China (NSFC grants 12141301, 12121003, 11825302). 
T.W. acknowledges the support from the National Natural Science Foundation of China (Project No. 12173017), and the China Manned Space Project with No. CMS-CSST-2021-A07. 
Y.S. thanks for the support by the New Cornerstone Science Foundation through the XPLORER PRIZE.  
GEM acknowledges the Villum Fonden research grant 13160 “Gas to stars, stars to dust: tracing star formation across cosmic time,” grant 37440, “The Hidden Cosmos,” and the Cosmic Dawn Center of Excellence funded by the Danish National Research Foundation under the grant No. 140. 
CGG acknowledges support from CNES.
ZJ  acknowledges funding from JWST/NIRCam contract to the University of Arizona NAS5-02015.
ID acknowledges support from INAF Minigrant "Harnessing the power of VLBA towards a census of AGN and star formation at high redshift".
CdE acknowledges funding from the MCIN/AEI (Spain) and the “NextGenerationEU”/PRTR (European Union) through the Juan de la Cierva-Formación program (FJC2021-047307-I).
SJ is supported by the European Union's Horizon Europe research and innovation program under the Marie Sk\l{}odowska-Curie grant agreement No. 101060888.
This research used APLpy, an open-source plotting package for Python \citep{aplpy2012, aplpy2019}.
This work is based on observations carried out under project number M21AA with the IRAM NOEMA Interferometer. IRAM is supported by INSU/CNRS (France), MPG (Germany) and IGN (Spain).
This research uses data obtained  through the Telescope Access
Program (TAP), which is funded by the National Astronomical Observatories, Chinese Academy of Sciences, and the Special Fund for Astronomy from the Ministry of Finance.
\end{acknowledgements}

\bibliographystyle{aa-yang} 
\bibliography{bibitem.bib} 
%

\appendix
\section{Supplementary data}
\begin{figure*}
\centering
\includegraphics[width=0.65\linewidth]{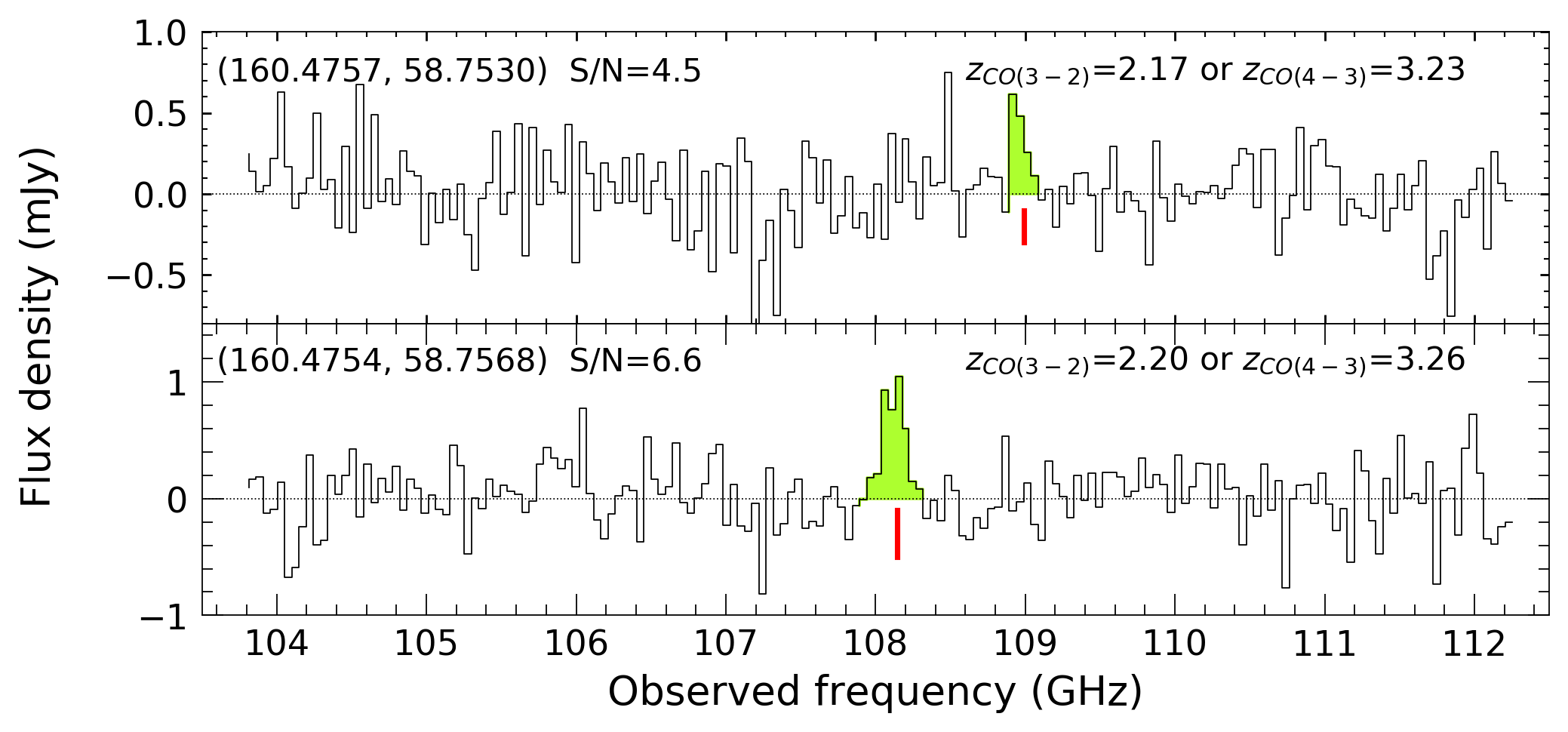}
\caption{Spectra of the two galaxies detected at 108GHz as discussed in Section~\ref{sec:lh-sbc3}.The emission lines are highlighted in yellow with the red vertical bars indicating the line centres. The coordinates, the signal-to-noise ratio of the  lines, and the implied redshifts are marked. }
\label{fig:foreground}
\end{figure*}


\end{document}